\documentclass[10pt,journal,compsoc]{IEEEtran}
\ifCLASSOPTIONcompsoc
\usepackage[nocompress]{cite}
\else
\usepackage{cite}
\fi

\ifCLASSINFOpdf
\else
\fi
\usepackage{amsmath,amssymb,amsfonts}
\usepackage{algorithm}
\usepackage{algorithmic}
\usepackage{graphicx}
\usepackage{textcomp}

\usepackage{booktabs} 
\usepackage{siunitx}
\usepackage{amsmath,xcolor,pifont}

\usepackage{color}
\usepackage{xcolor,colortbl}
\usepackage{multirow}
\usepackage{makecell}
\usepackage{tabularx}
\usepackage{listings}
\usepackage{graphicx}
\usepackage{subcaption}
\usepackage{enumitem}

\usepackage[colorinlistoftodos]{todonotes}
\newcommand{\red}[1]{\textcolor{red}{#1}}

\def\BibTeX{{\rm B\kern-.05em{\sc i\kern-.025em b}\kern-.08em
    T\kern-.1667em\lower.7ex\hbox{E}\kern-.125emX}}

\begin{document}
\newtheorem{defn}{Definition}


\title{Humanoid: A Deep Learning-based Approach to Automated Black-box Android App Testing}

\author{Yuanchun~Li,
        Ziyue~Yang,
        Yao~Guo,
        Xiangqun~Chen
\IEEEcompsocitemizethanks{\IEEEcompsocthanksitem Yuanchun Li, Ziyue Yang, Yao Guo, and Xiangqun Chen are with the Key Laboratory of High Confidence Software Technologies (Peking University), Ministry of Education, Beijing 100871, China.\protect\\
E-mail: {liyuanchun, ziyue.yang, yaoguo, cherry}@pku.edu.cn}
\IEEEcompsocitemizethanks{\IEEEcompsocthanksitem Yuanchun Li and Ziyue Yang contributed equally to this paper.}
}

\markboth{Technical Report}%
{Li \MakeLowercase{\textit{et al.}}: Humanoid: A Deep Learning-based Approach to Automated Black-box Android App Testing}

\IEEEtitleabstractindextext{%
\begin{abstract}
Automated input generators are widely used for large-scale dynamic analysis of mobile apps. Such input generators must constantly choose which UI element to interact with and how to interact with it, in order to achieve high coverage with a limited time budget. Currently, most input generators adopt pseudo-random or brute-force searching strategies, which may take very long to find the correct combination of inputs that can drive the app into new and important states. In this paper, we propose Humanoid, a deep learning-based approach to GUI test input generation by learning from human interactions. Our insight is that if we can learn from human-generated interaction traces, it is possible to automatically prioritize test inputs based on their importance as perceived by users.
We design and implement a deep neural network model to learn how end-users would interact with an app (specifically, which UI elements to interact with and how). Our experiments showed that the interaction model can successfully prioritize user-preferred inputs for any new UI (with a top-1 accuracy of 51.2\% and a top-10 accuracy of 85.2\%).

We implemented an input generator for Android apps based on the learned model and evaluated it on both open-source apps and market apps. The results indicated that Humanoid was able to achieve higher coverage than six state-of-the-art test generators. \red{However, further analysis showed that the learned model was not the main reason of coverage improvement. Although the learned interaction pattern could drive the app into some important GUI states with higher probabilities, it had limited effect on the width and depth of GUI state search, which is the key to improve test coverage in the long term. Whether and how human interaction patterns can be used to improve coverage is still an unknown and challenging problem.}
\end{abstract}




}



\maketitle

\IEEEdisplaynontitleabstractindextext
\IEEEpeerreviewmaketitle
\IEEEraisesectionheading{\section{Introduction}\label{sec:introduction}}

Mobile applications (\emph{apps} in short) have seen widespread adoption in recent years, with over three million apps available for download in both Google Play and Apple App Store, while billions of downloads have been accumulated~\cite{wiki:appstore, wiki:googleplay}. These apps need to be adequately inspected by the app markets and governments who want to prevent malicious or inappropriate apps being published. Many dynamic analysis methods \cite{enck2014taintdroid,dong2018frauddroid,hu2015protecting} are proposed for such purpose. These methods usually require running an app in a sandbox, and the completeness of analysis is based on how many functionalities in the app are covered. However, due to the huge amount of apps and limited human resources, it is difficult for auditors to manually run the apps. As a result, automated input generators for mobile apps have been studied extensively in both academia and industry.

The inputs for mobile apps are typically represented by the interactions with the graphical user interface (GUI) of an app. Specifically, an interaction may include clicking, scrolling, or inputting text into a GUI element, such as a button, an image, or a text block.
The job of an input generator is to produce a sequence of interactions for the app under analysis, which can be used to detect software-related problems, such as bugs, vulnerabilities, and security issues. The effectiveness of a input generator is often measured by its coverage.
Given unlimited time, one can potentially try all possible interaction sequences and combinations to achieve a perfect coverage. However, in real-world situations where the time budget for analysis is typically limited and the app may contain hundreds of GUI states and dozens of possible interactions in each state, an input generator can only choose a small subset of interaction sequences to explore.

The key to success for an automated input generator is to choose the ``best'' interactions for a given app, while the definition of ``best'' is based on the goal of analysis.
For example, if the input generator is used by developers to find bugs in their apps, the unexpected or corner-case input would be preferred. While if the purpose of analysis is to audit the apps (e.g. detecting inappropriate content, privacy leakage, accessibility issues, etc.), the best input would be the interactions that end-users are more likely to pick, so that the commonly-used content can get more attention during analysis.
Existing work on automated GUI input generation \cite{android:adb:monkey,Dynodroid:FSE:2013,GUIRipper:ASE:2012,GUICC:ASE:2016,Sapienz:ISSTA:2016,stoat:FSE:2017} are usually focused on the bug-finding scenario, where random strategies or corner-case-targeted strategies are effective.
However, these approaches may not be as effective in the dynamic analysis and auditing scenario, since they ignore the GUI information, thus is impossible to understand which interactions would be preferred by the users.
The key question we want to investigate in this paper is: \emph{Can we teach an automated input generator to prioritize inputs like a human being?}


\begin{figure}[tbp]
\centering
\includegraphics[width=3.5in]{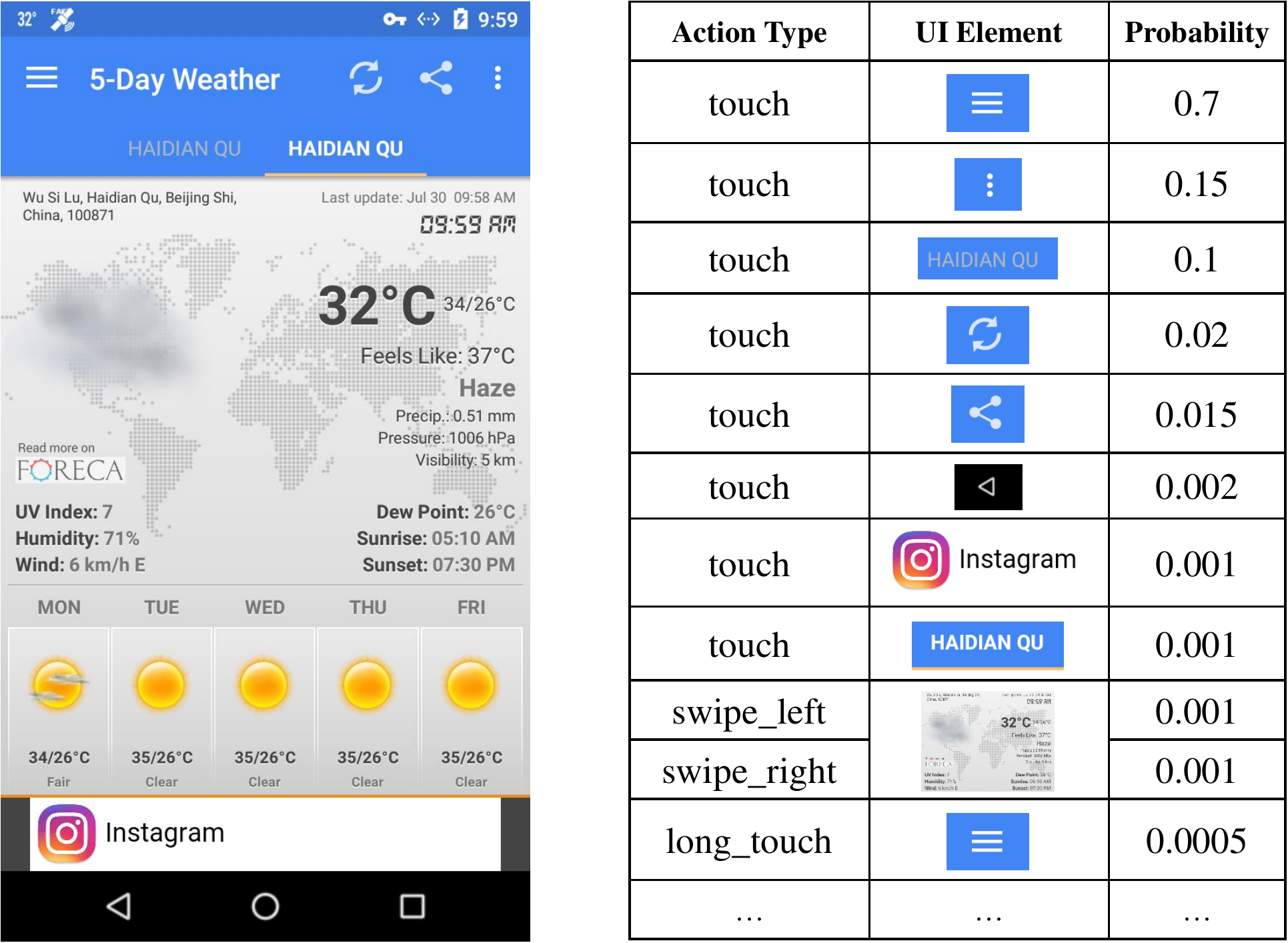}
\caption{An illustration of how Humanoid chooses inputs for a UI state. \small\emph{The left side is a screenshot of the current UI of an Android app under analysis, and the right side enumerates the most possible interactions in the UI state. Humanoid computes a probability for each action based on a model learned from human interaction traces. The probability represents how likely the action will be chosen by Humanoid as the test input.}}
\label{figure:running_example}
\end{figure}

This paper proposes Humanoid, an automated GUI input generator that is able to learn how humans interact with mobile apps and then use the learned model to guide input generation as a human. With the knowledge and model learned from human interaction traces, Humanoid can prioritize the possible interactions on a GUI page according to their importance from user perspective, thus generating inputs that can lead to important states faster and achieve higher coverage.

We can use the GUI page from an app shown in Figure~\ref{figure:running_example} as a motivating example. There are more than 20 actions that can be performed on that page, while most of them are ineffective or unrelated to the core functionalities of the app, such as swiping left on the current page, which is actually not scrollable, or clicking the advertisement on the bottom. While a random input generator may have to try all possible choices (including those ineffective ones), Humanoid is able to increase the probabilities of clicking the menu buttons, which are more likely to drive the app into additional important GUI states.

\textit{The core of Humanoid is a deep neural network (DNN) model that predicts which inputs are more likely to be generated by human users.} The input of the model is the current UI state as well as the most recent UI transitions, represented as a stack of images, while the output is a predicted distribution of possible next actions, including the action type and the corresponding location coordinates on the screen. By comparing the predicted distribution with all possible actions on the UI page, the model is able to assign a probability to each action.

We then design a biased random search algorithm to generate inputs based on the probabilities predicted by the DNN model. The algorithm always tries to pick an unexplored action in the current state as the input. If there are multiple unexplored actions, the one with the highest probability is picked. If all actions have already been explored, the algorithm will pick the action that can lead to the state with the most unexplored actions.
The reason why this strategy can improve coverage is that \emph{the important states that users prefer to visit and with more possible inputs are more frequently and adequately explored. Given a limited time budge, spending time on these important states can improve coverage more efficiently than on other random states.}


We implemented Humanoid and trained the interaction model with 304,976 human interactions extracted from a large-scale crowd-sourced UI interaction dataset Rico \cite{rico}. As a generic input generator, the model can be easily integrated with other dynamic analysis tools by simply replacing their input selection logic.

To evaluate the interaction model, we examined whether the model can learn human interaction patterns by using it to prioritize the possible actions for each UI state from a subset of the Rico interaction trace dataset. The results show that, for most UI states in the interaction traces, human-performed actions are ranked in the top 10\% across all actions according to Humanoid-predicted probabilities, which was significantly better than a random strategy whose expectation would be around 50\%.

To evaluate the input generation algorithm, we compared Humanoid with six state-of-the-art input generators in Android. These input generators are primarily designed for testing purpose (e.g. finding bugs), thus may not be very effective in covering apps' functionalities. The apps used for testing include 68 open-source apps obtained from the AndroTest \cite{androtest} dataset, which is a widely-used benchmark dataset for evaluating Android test input generators. We also tested 200 popular apps from Google Play\footnote{https://play.google.com/store/apps}, to see whether Humanoid is also effective for more complicated apps in real-world scenarios.
The results show that, Humanoid was able to achieve 43.3\% line coverage for open-source apps and 24.1\% activity coverage (the percentage of reached activities\footnote{https://developer.android.com/guide/components/activities}) for market apps, which was significantly higher than the best results (38.8\% and 19.7\%) achieved by other input generators within the same time duration. 

\red{We further analyzed Humanoid's input generation algorithm by replacing the learned interaction model with a random policy, in order to see whether the coverage improvement was actually brought by the learned model. The result showed that, given sufficient time, there was no significant difference between the test coverage achieved with and without the learned model. This is because that coverage improvement is primarily due to how many different states are visited, instead of how much or how fast a input generator can navigate into human-preferred states. The former mainly depends on the state space search algorithm, rather than the interaction model learned from human interaction traces.}

This paper makes the following main contributions:
\begin{enumerate}
\item To the best of our knowledge, this is the first work to introduce the idea of prioritizing GUI inputs by learning from human interaction traces, in order to generate effective inputs for automated dynamic mobile app analysis. 
\item We propose and implement a deep learning model to prioritize test inputs according to their importance from users' perspective. We demonstrate that Humanoid is able to rank the human-preferred interactions higher than others in new apps.
\item We design and implement a input generator, named Humanoid, to automatically generate interactions that are more frequently performed by users. We evaluate Humanoid with extensive experiments.
\end{enumerate}

\section{Background and Related Work}

\subsection{Android UI}
For a mobile app, \textit{user interface} (UI) is the place where interactions between humans and machines occur. App developers design UI to help users understand the features of their apps, and users interact with the apps through the UI. The graphical user interface (GUI) is the most important type of UI for most mobile apps, where apps present content and actionable widgets on the screen and users interact with the widgets using actions such as clicks, swipes, and text inputs.

The GUI pages (or screenshots) presented in mobile apps typically use a tree-structured layout. For example, in a screenshot of an Android app, all UI elements are built using View and ViewGroup objects and organized as a tree\footnote{https://developer.android.com/guide/topics/ui/declaring-layout}. A View is a leaf node that draws something on the screen that the user can see and interact with. A ViewGroup is a parent node that holds other nodes in order to define the layout of the interface. A UI state can be identified as a snapshot of the structure and content in the current UI tree, and a node in the UI tree is called a UI element.

An app can be viewed as a combination of many GUI states and the transitions between them. Each GUI state serves different functionalities or renders different content. App users navigate between UI states by interacting with the UI elements.

\subsection{Automated GUI Input Generation}

Automated GUI input generation has become an active research area since the prevalence of mobile apps. Most of the research work target at the Android platform, partly due to the popularity of Android apps, as well as the fragmentation of Android devices and OS releases.

In Android, input generators interact with apps in the same way as humans: sending simulated gestures to the GUI of an app. Since the acceptable gestures in a UI state are limited, the main difference between different test generators is their strategies used in prioritizing these actions. There are mainly three types of strategies: random, model-based, and targeted.

A typical example using the \textit{random strategy} is Monkey \cite{android:adb:monkey}, the official tool for automated app testing in Android. Monkey sends random types of input events to random locations on the screen without considering its GUI structure. DynoDroid \cite{Dynodroid:FSE:2013} also uses a random strategy, while the input sent by DynoDroid is smarter than Monkey: A lot of unacceptable events are filtered out based on the GUI structure and registered event listeners in an app. Sapienz \cite{Sapienz:ISSTA:2016} makes use of a genetic algorithm to optimize random test sequences. Polariz \cite{mao2017crowd} extracts and reuses ``motifs'' obtained by human testers to help generate random test sequences.

Several other testing tools build and use \textit{the GUI model} of mobile apps to generate test input. These models are usually represented as finite state machines that store the transitions between app window states. Such GUI models can be constructed dynamically \cite{li2017droidbot,PUMA:MobiSys:2014,DroidMate:MobileSoft:2016,stoat:FSE:2017,DroidMateM:MobileSoft:2018,GUIRipper:ASE:2012,GUICC:ASE:2016,SwiftHand:OOPSLA:2013,TestIndustrial:ASE} or statically \cite{gator:ASE:2015}. Based on the GUI models, testing tools can generate events that can quickly navigate the app to unexplored states. These model-based strategies can be further optimized in various ways. For example, Stoat \cite{stoat:FSE:2017} can iteratively refine the test strategy based on existing explorations, and DroidMate \cite{DroidMateM:MobileSoft:2018} can infer acceptable actions for a UI element by mining from other apps.

The \textit{targeted strategy} is designed to address the problem that some app behavior can only be revealed with specific test inputs. For example, a malicious app may only send SMS messages upon receiving a certain broadcast \cite{appcontext}. These testing tools \cite{ACTEve:FSE:2012,A3E:OOPSLA:2013,Brahmastra:SEC:2014,wong2016intellidroid} usually use sophisticated static analysis techniques such as data flow analysis and symbolic execution to find the interactions that can lead to the target states. However, their effectiveness can be easily affected by the complexity of app code and the difficulty of mapping code to UI elements.

Unlike existing input generators that are mainly designed for developers to test their own apps, our work aim to assist auditors (app markets and governments) in large-scale dynamic app analysis, such as inappropriate content detection \cite{hu2015protecting}, privacy leakage detection \cite{enck2014taintdroid}, ad fraud detection \cite{dong2018frauddroid}, etc. In such scenarios, the input generator should prioritize user-preferred interactions over unexpected or corner-case inputs, in order to trigger more functionalities that may be used by users.


\subsection{Software GUI Analysis}

GUI is an indispensable part of software on most major platforms including Android. Analyzing the app's GUI is of great interest to many researchers and practitioners. There are mainly two lines of research in this area. One is to understand the behavior of apps from the software engineering perspective. Another is from the human-computer interaction perspective to analyze the user interface design.

As mentioned before, many automated testing tools build and use GUI models to guide test input generation. Unlike such models that mainly use the transitions between UI states to abstract the app behavior, there are also some approaches focused on analyzing the information in each individual UI state. For example, Huang \textit{et al.} \cite{AsDroid:ICSE:2014} and Rubin \textit{et al.} \cite{rubin2015covert} proposed to detect stealthy behaviors in Android apps by comparing the actual behaviors with the UI. PERUIM \cite{peruim} extracted the mapping between an app's permissions to its UI to help users understand why each permission is requested, and AUDACIOUS \cite{audacious:2016} provided a way to control permission access based on UI components. Chen \textit{et al.} \cite{ui2code:ICSE:2018} introduced a machine learning-based method to extract UI skeletons from UI images, in order to facilitate GUI development.

In human-computer interaction research, software GUI is mainly used to mine UI design practices \cite{kumar2013webzeitgeist,alharbi2015designPattern} and interaction patterns \cite{deka2016erica} at scale. The mined knowledge can further be used to guide UI and UX (user experience) design. To facilitate mobile app design mining, Deka \textit{et al.} have collected and released a dataset named Rico \cite{rico}, which consists of a large number of UI screens and human interactions.

Our work lies in the intersection between software engineering and human-computer interaction: we propose a deep learning approach to mining human interaction patterns from the Rico dataset and use the learned patterns to guide automated input generation.

\section{Our Approach}

In order to employ human knowledge on mobile apps to augment mobile app testing, this paper proposes Humanoid, a new automated input generator that is able to prioritize inputs based on knowledge learned from human-generated app interaction traces. Similar to many existing input generators, Humanoid uses a GUI model to understand and explore the behavior of the app under analysis. However, unlike traditional model-based approaches that randomly choose an action to perform when exploring a UI state, Humanoid prioritize the actions that are more likely to be performed by human users. We expect that such strategy can drive the app into important states more frequently and achieve higher coverage than random strategies.

\subsection{Approach Overview}

\begin{figure*}[tbp]
\centering
\includegraphics[width=6.5in]{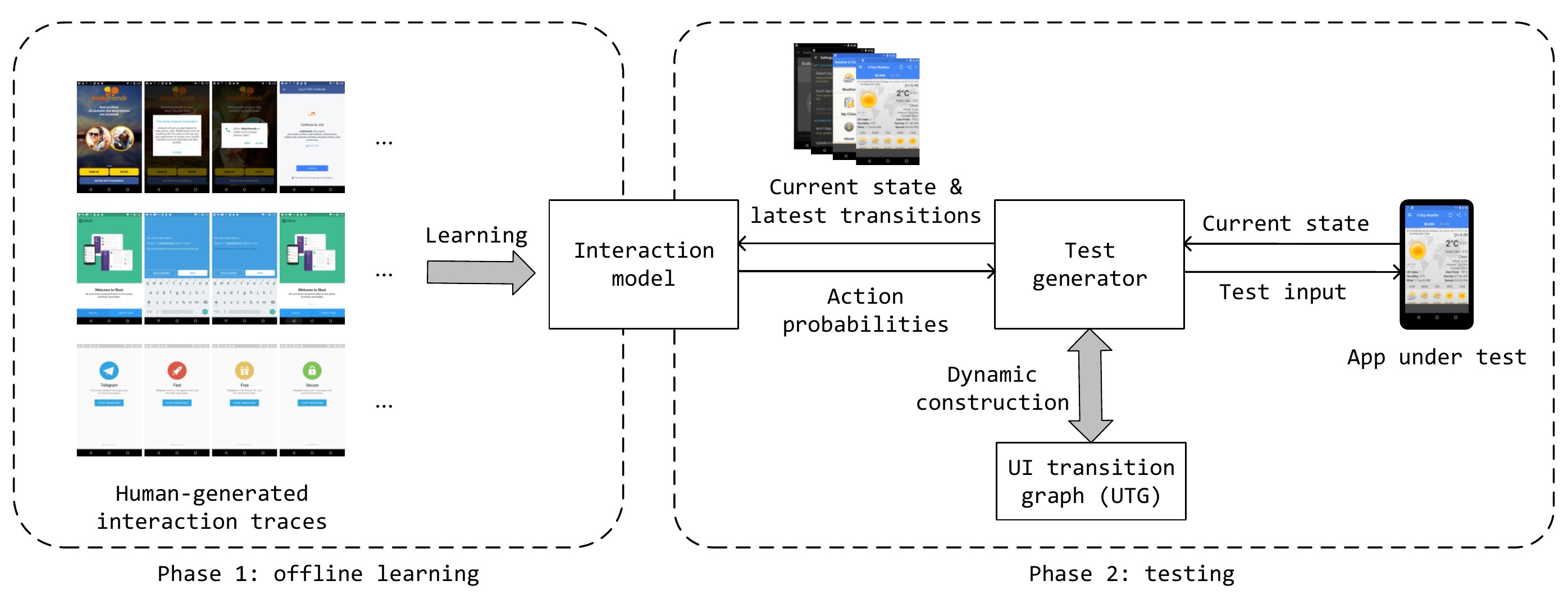}
\caption{Overview of our approach.}
\label{figure:overview}
\end{figure*}

Figure~\ref{figure:overview} shows the overview of Humanoid. The core of Humanoid is a deep learning model (i.e. interaction model) that learns the patterns about how humans interact with apps.
Based on the interaction model, the whole system can be separated into two phases, including \textit{an offline phase} for training the model with human-generated interaction traces and \textit{an online phase} in which the model is used to guide test input generation.

In the offline learning phase, we use a deep neural network model to learn the relation between the GUI contexts and user-performed interactions. A GUI context is represented as the visual information in the current UI state and the latest UI transitions, while an interaction is represented as the action type (touch, swipe, etc.) and the location coordinates of the action. After learning from large-scale human interaction traces, Humanoid is able to predict a probability distribution of the action type and action location for a new UI state. The predicted distribution can then be used to calculate the probability of each UI element being interacted with by humans and how to interact with it.

During the online testing phase, we adopt a biased random search algorithm to generate inputs. The algorithm maintains a GUI model named \textit{UI transition graph} (UTG) to remember the explored states and actions in the app. Both the GUI model and the interaction model are used by Humanoid to decide what input to send. The algorithm picks new actions to explore based on the probabilities predicted by the interaction model and navigate between explored UI states based on the UTG.

\subsection{Interaction Trace Preprocessing}

First of all, we will need a large dataset (Rico \cite{rico}) with human interaction traces to train a user interaction model, which is the key component in Humanoid. Because the human interaction traces in Rico are not designed for training for our purpose, we first need to preprocess the interaction traces.

A raw human interaction trace is usually a continuous stream of motion events sent to the screen \cite{rico}, where each motion event is comprised of when (the timestamp) and where (the x,y coordinate) the cursor (the user's finger) enters, moves, and leaves the screen.
The state change is also continuous because of the animations and dynamically loaded content.

The input acceptable to our model is a set of user interaction flows. Each interaction flow consists of a sequence of UI states $<s_1, s_2, s_3, ..., s_n>$ and a sequence of actions $<a_1, a_2, a_3, ..., a_n>$ that are taken in the corresponding UI states. To convert the raw interaction traces to the format acceptable to our model, we need to split cursor movements and identify user actions from them.

We consider seven types of user actions in Humanoid, including \texttt{touch}, \texttt{long\_touch}, \texttt{swipe\_ up/down/left/right}, and \texttt{input\_text}. Most use cases in apps can be accomplished with these types of actions. Each action is represented by the action type and the target location on the screen. 
In order to extract user actions from raw cursor traces, we first aggregate the cursor movements into \textit{interaction sessions}. 

An \textit{interaction session} is defined as the period between when the cursor enters the screen and when the cursor leaves the screen. We denote the timestamps of the session start and the session end as $time_{start}$ and $time_{end}$, and the cursor locations as $loc_{start}$ and $loc_{end}$.
Then we map interaction sessions to user actions according to a list of heuristic rules, as shown in Table~\ref{table:rules_extract_actions}.

\begin{table}[]
\caption{Rules to extract actions from cursor movements. Note that $|loc_{end}-loc_{start}|$ means the Euclidean distance between $loc_{start}$ and $loc_{end}$. The numbers are heuristically chosen to accommodate Android default configurations.}
\label{table:rules_extract_actions}
\begin{tabular}{p{1.7in}|p{0.7in}p{0.6in}}
\hline
\textbf{Condition}                                                 & \textbf{Action type} & \textbf{location} \\ \hline
$|loc_{end}-loc_{start}|<50px$ and $time_{end}-time_{start}<500ms$     & touch       & $loc_{start}$       \\ \hline
$|loc_{end}-loc_{start}|<50px$ and $time_{end}-time_{start}\geq 500ms$ & long\_touch & $loc_{start}$       \\ \hline
$|loc_{end}-loc_{start}| \geq 50px$ and $loc_{end}$ is on the left / right / top / bottom of $loc_{start}$                     & swipe\_left swipe\_right swipe\_up swipe\_down   & $loc_{start}$      \\ \hline
Successive interaction sessions where the keyboard is displayed and an editable element is focused & input\_text   & center of the editable element   \\ \hline
\end{tabular}
\end{table}

Once we have extracted the sequence of actions $<a_1, a_2, a_3, ..., a_n>$, we are able to match UI state changes with the actions based on the action timestamps. We use the UI state captured right before the timestamp of $a_i$ as $s_i$ to form the state sequence $<s_1, s_2, s_3, ..., s_n>$. The state sequence and the action sequence together represent a user interaction flow, which will be used as the training data for our human interaction model.

\subsection{Model Training}

This section explains in more details on how we use a deep neural network model to learn human interaction patterns from human interaction traces.

End-users interact with an app based on what they want to do with the app and what they see on its GUI. Since different apps often share common UI design patterns, it is intuitive that the way how humans interact with GUI is generalizable across different apps. The goal of the interaction model is to capture such generalizable interaction patterns.

\begin{figure}[tbp]
\centering
\includegraphics[width=3in]{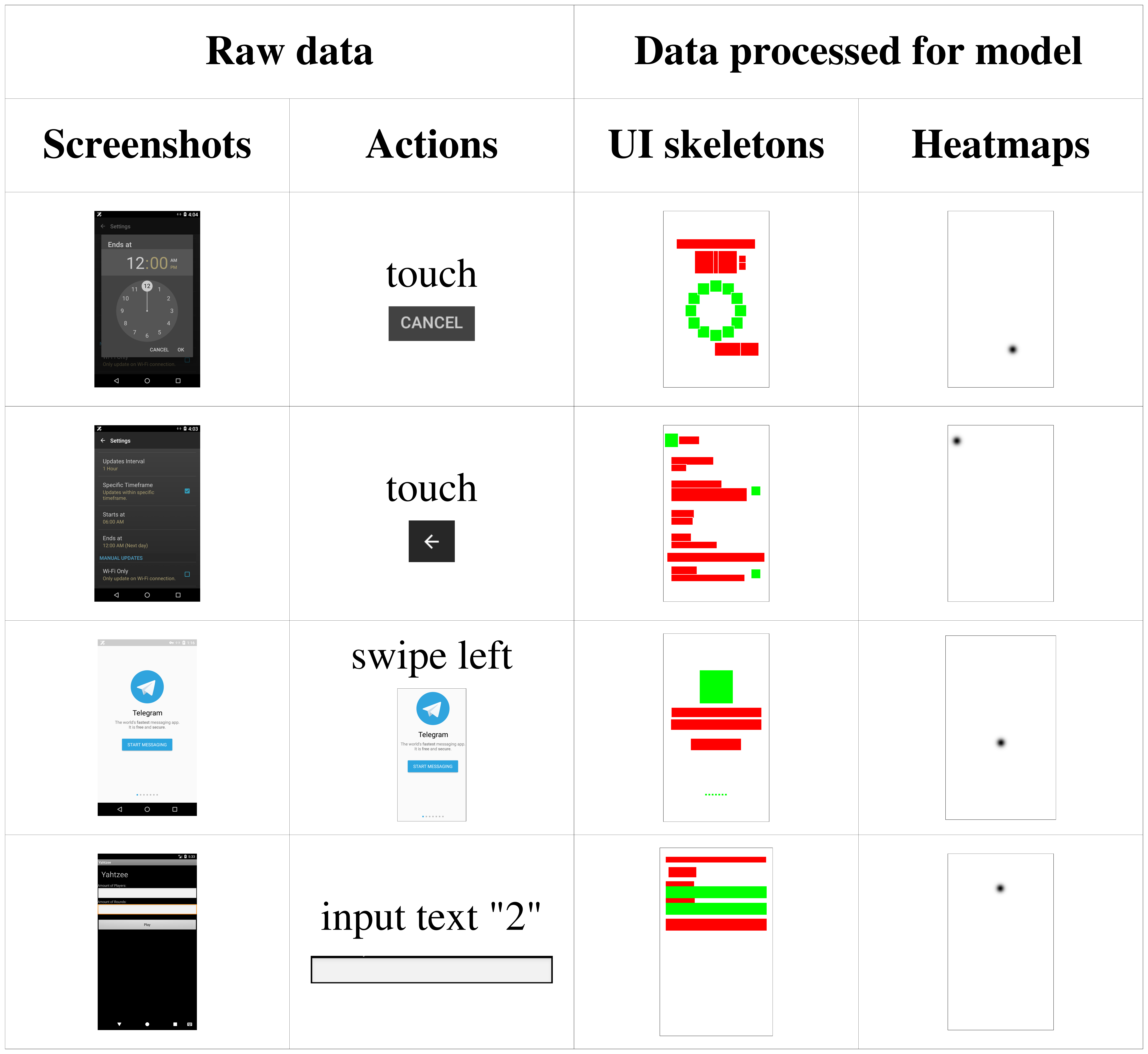}
\caption{The representation of UI states and actions in the interaction model.}
\label{figure:feature}
\end{figure}

We introduce a concept \emph{UI context} to model what humans reference when they interact with an app. A \emph{UI context} $context_i$  consists of the current UI state $s_i$ and three latest UI transitions $(s_{i-1},a_{i-1}),\ (s_{i-2},a_{i-2}),\ (s_{i-3},a_{i-3})$. The current UI state represents what the users see when they perform the action, while the latest UI transitions are used to model the users' underlying intention during the current interaction session. The reason why we used exactly three historic UI transitions is because most common interaction patterns contain no more than three actions \cite{deka2016erica}.

Figure~\ref{figure:feature} shows how we represent the UI states and actions in our model.
Each UI state is represented as a two-channel UI skeleton image, in which the first channel (red channel) renders the bounding box regions of text UI elements and the second channel (green channel) renders the bounding box regions of non-text UI elements. The reason why we use the UI skeletons instead of the original screenshots is that most characters on the screenshots do not affect how humans interact with the apps. For example, the UI style (font size, button style, background color, etc.) of the same app may change across different OS and app versions, whereas the way how users use the app remains the same. Some apps even provide functionalities like ``night mode'' to allow users change the UI styles internally. Such UI style characters may bring noises to our model and affect the model's generalization ability, thus we exclude them from the input representation.

Each action is represented by its action type and target location coordinates. The action type is encoded as a seven-dimensional vector, in which each dimension maps to one of the seven action types as described earlier. The action target location is encoded as a \textit{heatmap}. Each pixel in the heatmap is the probability of the pixel being the action target location. We use a heatmap rather than the raw coordinates to represent the action location because the raw coordinates are highly non-linear and more difficult to learn \cite{tompson2014joint}.


In summary, the representation of a \emph{UI context}, i.e. the input feature for our interaction model, is a stack of images including one 2-channel image for the current UI state and three 3-channel images for three latest UI transitions (each transition include one 2-channel image for the UI state and one 1-channel image for the action). All images are scaled to the size of \texttt{180x320} pixels. For ease of learning, we also add one channel of zero padding for the current UI state. In the end, a \emph{UI context} is represented as a \texttt{4x180x320x3} vector.

Given the \emph{UI context} vector, the output of the interaction model is an ``action'' that is likely to be performed by humans in the current state. Note that the predicted ``action'' is not an actual acceptable action in the current UI state. Instead, it is a probability distribution of types and locations of the expected human-like actions. Specifically, the goal of the model is to learn two conditional probability distributions:

\begin{enumerate}
\item $p_{type}(\ t\ |\ context_i)$

where $t \in \{touch, long\_touch, swipe\_up, ...\}$, meaning the probability distribution of $t$, the type of the next action $a_{i}$, given the current \emph{UI context}.
\item $p_{loc}(\ x,y\ |\ context_i)$

where $0 < x < screen\_width$ and $0 < y < screen\_height$, meaning the probability distribution of the target location $x,y$ of the next action $a_{i}$, given the current \emph{UI context}.
\end{enumerate}

\begin{figure*}[tbp]
\centering
\includegraphics[width=6in]{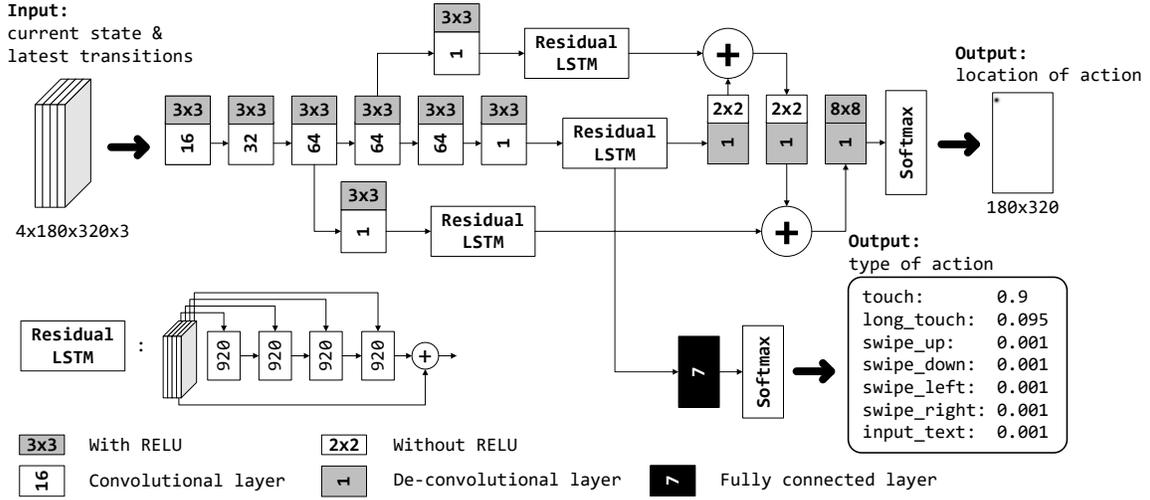}
\caption{The architecture of the deep learning model used in Humanoid. \emph{The model takes the representation of the current \emph{UI context} (current UI state and latest UI transitions) as input, and outputs a probability distribution of the next action (including the action type and the action location).}}
\label{figure:model}
\end{figure*}

Figure~\ref{figure:model} shows the deep neural network model used to learn the two conditional probability distributions defined above. It accepts the representation of the current \emph{UI context} $context_i$ as input, and outputs location and type distributions of $a_{i}$. The model consists of five main components: convolutional layers, residual LSTM modules, de-convolutional layers, a fully connected layer and loss functions.

\textbf{Convolutional layers.} Convolutional network structure has become a popular approach for image feature extraction, since it has been proved very powerful in computer vision tasks on large real-world datasets \cite{krizhevsky2012imagenet}. In our model, we use 5 convolutional layers with ReLU activations to extract features from UI skeleton images and action heatmaps. After each convolutional layer, there is a stride-2 max-pooling layer that reduces the width and height of its input to half. The pooling layers also help the model to identify UI elements having the same shape but different surroundings.

\textbf{Residual LSTM modules.} LSTM (Long-Short-Term Memory) networks are widely adopted in sequence modeling problems, such as machine translation \cite{sutskever2014sequence}, video classification \cite{donahue2015long}, etc. In our model, extracting features from historical transitions is also a sequence modeling problem. 
We insert residual LSTM modules after each of the last 3 convolutional layers, in order to capture UI transition sequence features on different resolution levels. In a residual LSTM module, the last dimension of input and the output of the normal LSTM are directly added through a residual path. 

Such residual structure makes the neural network easier to optimize \cite{he2016deep}, and gives hint that the location of an action should lie inside a UI element. To decrease model complexity, we also added a \texttt{1x1} convolutional layer before each residual LSTM module to reduce the feature dimension.

\textbf{De-convolutional layers.} This component is used to generate high-resolution probability distributions from the low-resolution output of residual LSTM modules. There are several options to accomplish this, such as bilinear interpolation, de-convolution, etc. We use de-convolutional layers, as it is easier to integrate with deep neural networks and more general than the interpolation methods. Features on different resolution levels are combined to improve the quality of generated heatmap \cite{long2015fully}. A softmax layer is followed to normalize the generated heatmap so that all pixels in the heatmap sum to 1, which is the probability distribution of action locations.

\textbf{Fully connected layer.} A single fully connected layer with softmax is used to generate the probability distribution of action types.

\textbf{Loss functions.} The model predicts both the action location and action type as probability distributions. Thus their cross-entropy losses against the ground truth (the action performed by humans) are suitable for model optimization. We use the sum of these two losses and a layer weight regularizer (to avoid overfitting) as the final loss function in the training process.

During training, each action $a_i$ in an interaction flow ($<s_1, s_2, s_3, ..., s_n>$, $<a_1, a_2, a_3, ..., a_n>$) is converted to the following probability distributions:
$$p_{type}(t) = 
\begin{cases}
1& t = a_i.type\\
0& \text{otherwise}
\end{cases}$$ and 
$$p_{loc}(x,y) = f(x - a_i.x,\ y - a_i.y)$$
where $f$ is the density function of the Gaussian distribution with $variance = 20$. We use the Gaussian distribution to approximate the probability distribution of actual screen coordinates recognized by a device, when the same UI element is interacted by many people for many times.

Similarly, when applying the model, we feed it with the representation of the current \emph{UI state} to predict the probability distributions $p_{type}(t)$ and $p_{loc}(x,y)$ for the next action. As the predicted distributions cannot be directly used to guide test generation, we need to further convert them to the probabilities of the actions that can be performed in the current state. In order to do that, we first traverse the UI tree to find all possible actions in the current state, with each action containing the action type (denoted as $action.type$) and the action target element (denoted as $action.element$). Then we calculate the probability of each action based on the distribution predicted by the model:
$$p(action) = p_{type}(action.type) * \sum_{x,y\text{ in action.element}} p_{loc}(x,y)$$
The action probabilities can finally be used to guide test input generation in the next step.

\subsection{Test Input Generation with Humanoid}

In this section, we describe how we apply the learned human interaction model to generate human-like test inputs.

Humanoid generates two types of test inputs, including explorations and navigations. \textit{Exploration inputs} are used to discover the unseen behaviors in an app, while \textit{navigation inputs} drive the app to known states that contain unexplored actions. When choosing from exploration inputs, the test generator does not know about the consequences of each test input, and the decision is made based on the guidance of the human interaction model (traditional test generators usually choose the input randomly). When generating navigation inputs, the test generator knows the target states of the input, as it has saved the memory of the transitions.

\begin{figure}[tbp]
\centering
\includegraphics[width=3in]{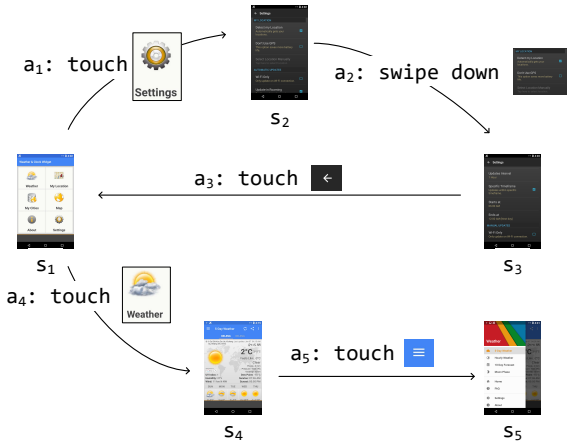}
\caption{An example of a UI transition graph (UTG).}
\label{figure:utg}
\end{figure}

Similar to many existing test generators, Humanoid uses a GUI model to save the memory of transitions. The GUI model we use is represented as \textit{UI transition graphs} (UTG in short), which is a directed graph, whose nodes are UI states and edges are the actions that lead to UI state transitions. The UTG is constructed at runtime: each time the test generator observes a new state $s_i$, it adds a new edge $<s_{i-1}, a_{i-i}, s_i>$ to the UTG, where $s_{i-1}$ is the last observed UI state and $a_{i-i}$ is the action performed in $s_{i-1}$. Figure~\ref{figure:utg} shows an example of UTG. With the UTG, the test generator can navigate to any known state by following the path to the state.

To decide between exploration and navigation and generate the input action, Humanoid adopts a simple but effective strategy, which is shown in Algorithm~\ref{algorithm:strategy}. In each step, Humanoid checks whether there are unexplored actions in the current state. Humanoid chooses exploration if there are unexplored actions (line 8), and chooses navigation if the current state is fully explored (line 10 to 12). The navigation process is straightforward. In the exploration process, Humanoid gets the probabilities of the actions predicted by the interaction model, and makes a weighted choice based on the probabilities. Since the actions that humans would take will be assigned higher probabilities, they get higher chances to be chosen by Humanoid as test inputs. Thus the inputs generated by Humanoid are more human-like than randomly chosen ones.

\begin{algorithm}[h!]
\caption{The biased random search strategy used by Humanoid to generate inputs.}
\label{algorithm:strategy}
\begin{algorithmic}[1]
\STATE Load the pretrained interaction model $M$
\STATE Create an empty UTG $G = <S, E>$
\STATE Start the app under test
\STATE Observe current UI state $s$ and add $s$ to $S$
\REPEAT
  \STATE Extract all actions in $s$ as $A$, $A_1 \subset A$ is the set of unexplored actions, $A_2 \subset A$ is the set of explored actions.
  \IF{$A_1$ is not empty}
    \STATE Choose action $a \in A$ with the highest probabilities predicted by $M$
  \ELSE
    \STATE Get $s' \in S$ that has the most unexplored actions
    \STATE Get the shortest path $p$ from $s$ to $s'$ in $G$
    \STATE Choose the first action in $p$ as $a$
  \ENDIF
  \STATE Perform action $a$
  \STATE Observe the new UI state $s_{new}$ and add $s_{new}$ to $S$
  \STATE Add the edge $<s, a, s_{new}>$ to $E$
\UNTIL{all actions in all states in $S$ have been explored}
\end{algorithmic}
\end{algorithm}

Compared to the existing testing tools, the main feature of Humanoid (and the main difference between different model-based test generators) is how the exploration input is chosen (line 8). Humanoid prioritizes the more valuable actions in exploration based on the interaction model, which has been trained from human interaction traces. This feature makes it faster to discover the correct input sequences, which in turn will drive the app into important UI states, thus leading to higher test coverage.

\section{Evaluation}

We evaluate Humanoid by primarily looking at the following aspects:
\begin{enumerate}
    \item Can Humanoid's deep learning model learn and predict correct human interaction patterns? What is its prediction accuracy? What is the learning and predicting cost? (Section~\ref{eval:model})
    \item Can Humanoid generate effective inputs for new Android apps? How does it compare with existing testing tools? (Section~\ref{eval:coverage})
    \item How effective is the interaction model learned from human traces? Is the high coverage brought by the interaction model? (Section~\ref{eval:effectiveness})
\end{enumerate}


\subsection{Experimental Setup}

\begin{figure}[tbp]
\centering
\includegraphics[width=2.8in]{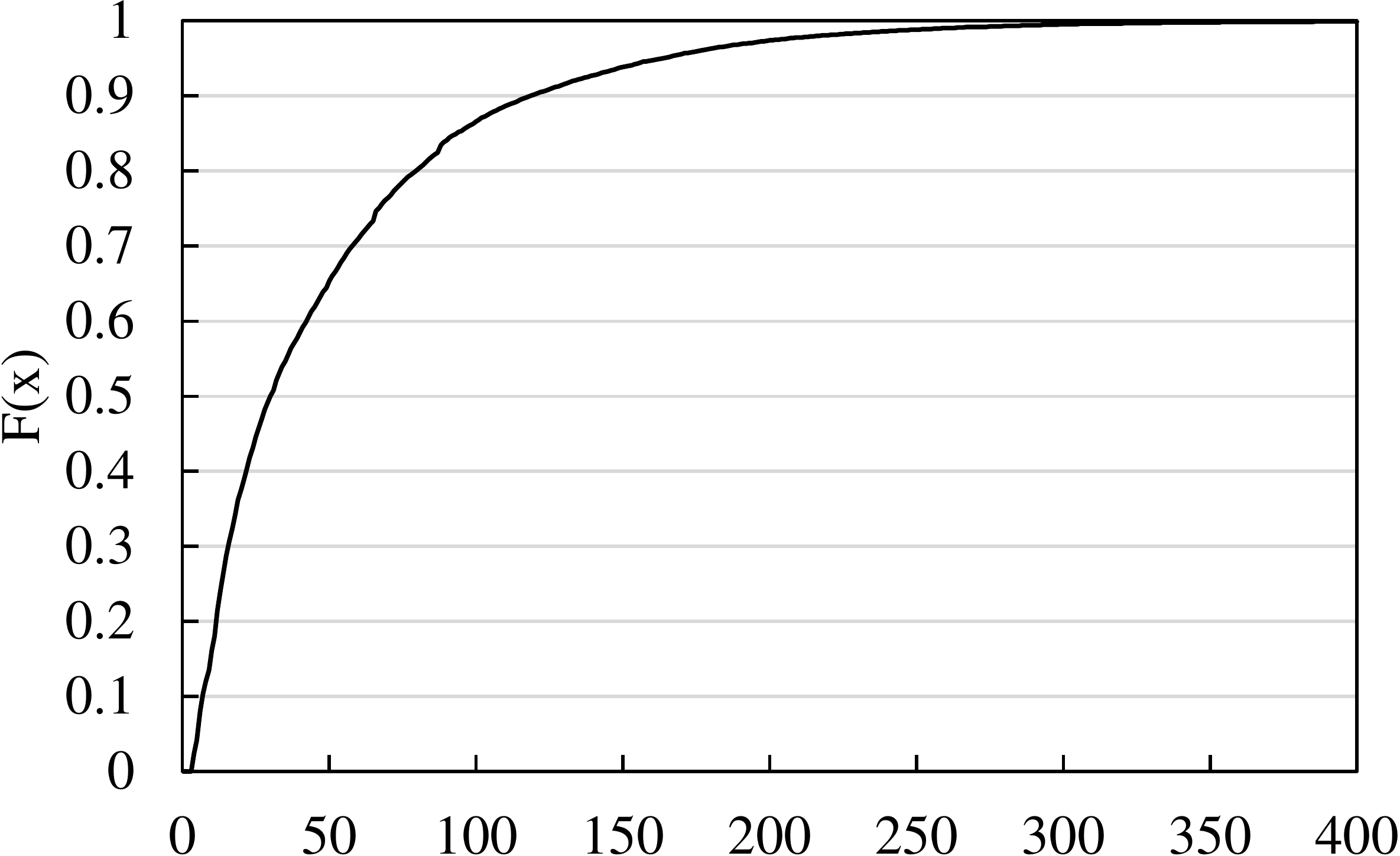}
\caption{CDF for the number of actions for each UI state.}
\label{figure:num_action_cdf}
\end{figure}

The dataset we used to train the Humanoid model is processed from Rico \cite{rico}, a large crowd-sourced dataset of human interactions. We extracted interaction flows from the raw data by identifying action sequences and state sequences. In the end, we obtained 12,278 interaction flows belonging to 10,477 apps. Each interaction flow contains 24.8 states on average. The cumulative distribution function (CDF) for the number of possible actions in each UI state is shown in Figure~\ref{figure:num_action_cdf}. On average, each UI state has 50.7 possible action candidates, while more than 10\% of the UI states include more than 100 possible action candidates.

The machine we used to train and test the interaction model is a workstation with two Intel Xeon E5-2620 CPUs, 64GB RAM and an NVidia GeForce GTX 1080 Ti GPU. The operating system of the machine was Ubuntu 16.04. The model was implemented with Tensorflow \cite{abadi2016tensorflow}.

For experiments on testing real Android apps, we used 4 computers with the same hardware and software as above, and ran 4 instances of Android emulators on each machine to test apps in parallel. The Android apps we used during testing evaluation include 68 open-source apps obtained from AndroTest \cite{androtest}, a commonly used dataset for evaluating Android test input generators, and 200 popular commercial apps downloaded from Google Play. We used Emma \cite{emma:java:coverage} to measure line coverage when testing the open-source apps. For the market apps without source code available, we used activity coverage (the percentage of reached activities, an activity for an Android app is like a webpage for a website) instead to measure the testing performance.


\subsection{Evaluation of the Deep Learning Model}
\label{eval:model}

In this experiment, we trained and tested our interaction model on existing human interaction traces to evaluate whether the model is able to learn how humans interact with apps. 

\subsubsection{Model Accuracy}

We randomly selected 100 apps from the dataset and used their interaction traces for testing. The interaction traces for the remaining 10,377 apps were used for training. In total, we had 302,382 UI states for training and 2,594 UI states for testing. For each UI state in the testing set, we used the interaction model to predict the probabilities for all possible actions and sort the actions in the descending order of predicted probabilities. The actions performed by humans were considered as the ground truth.

\begin{table}[tbp]
\centering
\caption{Top-N accuracy of the Humanoid interaction model in prioritizing human-generated actions, as compared with the random strategy.}
\label{table:top_n_accuracy}
\begin{tabular}{lrr}
\hline
\textbf{N} & \textbf{Random top-N accuracy} & \textbf{Humanoid top-N accuracy} \\ \hline
1  &  5.8\% & 51.2\% \\
3  & 17.5\% & 67.6\% \\ 
5  & 29.1\% & 74.8\% \\ 
10 & 58.3\% & 85.2\% \\ \hline
\end{tabular}
\end{table}

Table~\ref{table:top_n_accuracy} shows the accuracy of the Humanoid interaction model in prioritizing the human-performed actions in each UI state. Specifically, we calculated the probability that the ground truth (the human-performed action) ranks within top N (N=1,3,5,10) in the order of actions predicted by the interaction model. For comparison, we also calculated the top-N accuracies for the random strategy, i.e. the probability that the ground truth ranks top N if the actions are in random order. According to the results, our interaction model can identify and prioritize the human-generated actions with a higher accuracy. 
In particular, Humanoid was able to assign the highest probability to the human-generated action for more than 50\% of the UI states. We also calculated the percentile rank of the human-generated action in each UI state. The mean percentile rank was 20.6\% and the median was 9.5\%, meaning that Humanoid was able to prioritize the human-like actions into the top 10\% for most UI states.

\subsubsection{Training and Prediction Cost}

We then evaluated the cost of the interaction model. It took about 66 hours to train the interaction model with the dataset that contains 304,976 human-generated actions. It is acceptable since the model only needs to be trained once before being used for testing.
The average time spent to predict the probabilities of actions for a UI state was 107.9 milliseconds. Given the fact that it typically takes more than 2 seconds for an Android test generator to send a test input and wait for the new page being loaded, the time overhead that our interaction model would bring to the test generator is minimal.

\subsection{Test Coverage}
\label{eval:coverage}

In this experiment, we used Humanoid to generate test inputs for Android apps. We evaluated the test generator by examining whether it can actually improve test coverage. 

\subsubsection{Testing on Open-source Apps}

We test Humanoid with 68 open-source apps (none of them is in the training set of the interaction model) and compare it with six state-of-the-art test generators for Android, including Monkey \cite{android:adb:monkey}, PUMA \cite{PUMA:MobiSys:2014}, Stoat \cite{stoat:FSE:2017}, DroidMate \cite{DroidMateM:MobileSoft:2018}, Sapienz\cite{Sapienz:ISSTA:2016} and DroidBot \cite{li2017droidbot}. All tools were used with their default configurations. The input speeds of PUMA, Stoat, DroidMate and DroidBot were close to \textit{600 events/hour}, as they all need to read the UI state before performing an action and wait for the state transition after sending input, while Monkey and Sapienz could send input events with a very high speed (about \textit{6000 events/hour} in our experiments). 

We used each testing tool to run each open-source app for one hour. In order to accommodate the recent market apps, most of the tools were evaluated on Android 6.0, as it was supported by most of the tools (some with minor modification). However, as Sapienz is close-sourced and only supports Android 4.4, so it was evaluated on Android 4.4 instead. For each app and tool, we recorded the final coverage and the progressive coverage after each action was performed. We repeated this process three times and used the average as the final results.

\begin{table}[tbp]
\centering
\caption{Line coverage achieved by each testing tool for each open-source app. MO, PU, ST, DM, SA, DB, HU are short for Monkey, PUMA, Stoat, DroidMate, Sapienz, DroidBot, and Humanoid respectively. ``-'' means that the tool crashed during testing.}
\label{table:line_coverage_detailed}
\resizebox{0.48      \textwidth}{!}{%
\begin{tabular}{cl|rrrrrrr}
\hline
\textbf{ID} & \textbf{App package name} & \textbf{MO} & \textbf{PU} & \textbf{ST} & \textbf{DM} & \textbf{SA} & \textbf{DB} & \textbf{HU} \\
\hline
1 & com.example.amazed & \cellcolor{blue!20}  67\% & 50\% & 57\% & 14\% & 60\% & 63\% & \cellcolor{blue!20}  67\% \\ 
2 & com.example.anycut & 61\% & 61\% & 11\% & 48\% & 59\% & 62\% & \cellcolor{blue!20}  63\% \\ 
3 & com.google.android.divideandconquer & \cellcolor{blue!20}  80\% & - & 45\% & 45\% & 71\% & 52\% & 58\% \\ 
4 & com.android.lolcat & 24\% & - & 23\% & 17\% & 22\% & 23\% & \cellcolor{blue!20}  27\% \\ 
5 & info.bpace.munchlife & 58\% & 44\% & 40\% & 44\% & 56\% & 41\% & \cellcolor{blue!20}  65\% \\ 
6 & org.passwordmaker.android & 47\% & 29\% & 35\% & 39\% & 46\% & 43\% & \cellcolor{blue!20}  58\% \\ 
7 & com.google.android.photostream & 19\% & 22\% & 24\% & 12\% & \cellcolor{blue!20} 27\% & 22\% & \cellcolor{blue!20}  27\% \\ 
8 & com.bwx.bequick & 40\% & 33\% & 30\% & 35\% & 38\% & 37\% & \cellcolor{blue!20}  46\% \\ 
9 & com.example.android.musicplayer & \cellcolor{blue!20}  53\% & 51\% & 42\% & 52\% & 52\% & 52\% & \cellcolor{blue!20}  53\% \\ 
10 & com.google.android.opengles.spritetext & \cellcolor{blue!20}  61\% & 58\% & 60\% & 2\% & 58\% & 59\% & 59\% \\ 
11 & com.nloko.android.syncmypix & 20\% & - & 11\% & 4\% & 19\% & 20\% & \cellcolor{blue!20}  21\% \\ 
12 & com.google.android.opengles.triangle & \cellcolor{blue!20}  76\% & 72\% & \cellcolor{blue!20}  76\% & 19\% & 72\% & 75\% & 75\% \\ 
13 & a2dp.Vol & 33\% & 25\% & 19\% & 16\% & 35\% & 32\% & \cellcolor{blue!20}  39\% \\ 
14 & org.jtb.alogcat & 60\% & 64\% & 57\% & \cellcolor{blue!20}  66\% & 63\% & 65\% & 62\% \\ 
15 & aarddict.android & \cellcolor{blue!20}  13\% & 12\% & 12\% & 12\% & \cellcolor{blue!20} 13\% & 12\% & \cellcolor{blue!20}  13\% \\ 
16 & net.sf.andbatdog.batterydog & \cellcolor{blue!20}  64\% & 15\% & 29\% & 54\% & 61\% & 48\% & 62\% \\ 
17 & be.ppareit.swiftp\_free & 15\% & - & 13\% & 14\% & 12\% & 16\% & \cellcolor{blue!20}  17\% \\ 
18 & caldwell.ben.bites & 30\% & 15\% & 19\% & 16\% & 28\% & 16\% & \cellcolor{blue!20}  36\% \\ 
19 & ch.blinkenlights.battery & 74\% & 42\% & 50\% & 13\% & 72\% & 45\% & \cellcolor{blue!20}  75\% \\ 
20 & com.addi & \cellcolor{blue!20}  18\% & 17\% & 16\% & 17\% & 18\% & 17\% & \cellcolor{blue!20}  18\% \\ 
21 & com.angrydoughnuts.android.alarmclock & 64\% & - & 36\% & 41\% & 65\% & 61\% & \cellcolor{blue!20}  66\% \\ 
22 & com.chmod0.manpages & 42\% & 47\% & 50\% & \cellcolor{blue!20}  57\% & - & 52\% & 42\% \\ 
23 & com.everysoft.autoanswer & \cellcolor{blue!20}  14\% & 13\% & 11\% & 10\% & 13\% & 13\% & 13\% \\ 
24 & com.gluegadget.hndroid & \cellcolor{blue!20}  15\% & 9\% & 7\% & 3\% & 9\% & 6\% & 9\% \\ 
25 & com.hectorone.multismssender & 42\% & 23\% & 25\% & 24\% & \cellcolor{blue!20} 55\% & 26\% & 51\% \\ 
26 & com.irahul.worldclock & 90\% & 85\% & 86\% & 87\% & 87\% & 87\% & \cellcolor{blue!20}  93\% \\ 
27 & com.kvance.Nectroid & 32\% & 27\% & 29\% & 27\% & 42\% & 27\% & \cellcolor{blue!20}  53\% \\ 
28 & com.morphoss.acal & 14\% & - & 13\% & - & \cellcolor{blue!20} 23\% & 16\% & 22\% \\ 
29 & com.teleca.jamendo & 37\% & 42\% & 14\% & 3\% & 21\% & 22\% & \cellcolor{blue!20}  43\% \\ 
30 & com.templaro.opsiz.aka & 53\% & 42\% & 46\% & 61\% & - & 54\% & \cellcolor{blue!20}  82\% \\ 
31 & com.tum.yahtzee & 44\% & 8\% & 10\% & 7\% & 8\% & 10\% & \cellcolor{blue!20}  59\% \\ 
32 & com.zoffcc.applications.aagtl & 15\% & - & 15\% & 12\% & 16\% & 14\% & \cellcolor{blue!20}  19\% \\ 
33 & de.homac.Mirrored & 38\% & 29\% & 36\% & 8\% & 29\% & 22\% & \cellcolor{blue!20}  59\% \\ 
34 & org.dnaq.dialer2 & 36\% & 31\% & 37\% & 35\% & 38\% & 39\% & \cellcolor{blue!20}  40\% \\ 
35 & edu.killerud.fileexplorer & 24\% & 31\% & 20\% & 14\% & \cellcolor{blue!20} 40\% & 14\% & 16\% \\ 
36 & demo.killerud.gestures & \cellcolor{blue!20}  36\% & 22\% & 28\% & 22\% & 22\% & 32\% & 32\% \\ 
37 & com.smorgasbork.hotdeath & \cellcolor{blue!20}  62\% & 41\% & 29\% & 48\% & 53\% & 62\% & 54\% \\ 
38 & hu.vsza.adsdroid & \cellcolor{blue!20}  31\% & 28\% & 27\% & 24\% & 29\% & \cellcolor{blue!20}  31\% & 29\% \\ 
39 & i4nc4mp.myLock & \cellcolor{blue!20}  29\% & 23\% & 28\% & 26\% & 25\% & 27\% & 27\% \\ 
40 & in.shick.lockpatterngenerator & 69\% & 52\% & 46\% & 57\% & 71\% & 56\% & \cellcolor{blue!20}  76\% \\ 
41 & jp.sblo.pandora.aGrep & 40\% & 25\% & 20\% & 23\% & 45\% & 26\% & \cellcolor{blue!20}  47\% \\ 
42 & com.fsck.k9 & 4\% & 3\% & 3\% & 3\% & 4\% & 3\% & \cellcolor{blue!20} 6\% \\ 
43 & net.jaqpot.netcounter & 19\% & 33\% & 18\% & 16\% & \cellcolor{blue!20} 43\% & 16\% & 22\% \\ 
44 & org.beide.bomber & \cellcolor{blue!20}  73\% & 47\% & 55\% & 24\% & 64\% & 50\% & 68\% \\ 
45 & org.jfedor.frozenbubble & \cellcolor{blue!20}  71\% & 40\% & 50\% & 9\% & 52\% & 52\% & 61\% \\ 
46 & org.liberty.android.fantastischmemo & 19\% & 22\% & 15\% & 4\% & 27\% & 30\% & \cellcolor{blue!20}  31\% \\ 
47 & org.scoutant.blokish & \cellcolor{blue!20}  51\% & 32\% & 35\% & 21\% & 42\% & 44\% & 46\% \\ 
48 & org.smerty.zooborns & 20\% & 15\% & 23\% & 16\% & 14\% & 19\% & \cellcolor{blue!20}  24\% \\ 
49 & org.waxworlds.edam.importcontacts & 32\% & 4\% & 6\% & 38\% & 23\% & 36\% & \cellcolor{blue!20}  41\% \\ 
50 & org.wikipedia & 23\% & 20\% & 23\% & 11\% & 29\% & 20\% & \cellcolor{blue!20} 30\% \\ 
51 & com.android.keepass & 6\% & 2\% & 4\% & 4\% & 7\% & 5\% & \cellcolor{blue!20}  9\% \\ 
52 & hiof.enigma.android.soundboard & \cellcolor{blue!20}  42\% & 31\% & 31\% & \cellcolor{blue!20}  42\% & 31\% & \cellcolor{blue!20}  42\% & \cellcolor{blue!20}  42\% \\ 
53 & net.everythingandroid.timer & 73\% & 51\% & 50\% & 49\% & 58\% & 54\% & \cellcolor{blue!20}  75\% \\ 
54 & com.ringdroid & 16\% & \cellcolor{blue!20}  21\% & 1\% & 17\% & 15\% & 19\% & 17\% \\ 
55 & com.android.spritemethodtest & 54\% & 34\% & 28\% & 67\% & 79\% & 60\% & \cellcolor{blue!20}  81\% \\ 
56 & com.beust.android.translate & 41\% & 29\% & 31\% & 27\% & 44\% & 29\% & \cellcolor{blue!20}  48\% \\ 
57 & com.eleybourn.bookcatalogue & 19\% & 3\% & 4\% & 3\% & 10\% & 11\% & \cellcolor{blue!20}  20\% \\ 
58 & org.tomdroid & 42\% & 38\% & 32\% & 18\% & 40\% & 38\% & \cellcolor{blue!20}  49\% \\ 
59 & org.wordpress.android & 4\% & 4\% & 3\% & 3\% & 3\% & 4\% & \cellcolor{blue!20} 5\% \\ 
60 & com.evancharlton.mileage & 21\% & - & 15\% & 27\% & \cellcolor{blue!20} 34\% & 30\% & 29\% \\ 
61 & cri.sanity & 16\% & 11\% & 12\% & 14\% & 15\% & 16\% & \cellcolor{blue!20}  19\% \\ 
62 & org.jessies.dalvikexplorer & 17\% & 46\% & 33\% & 50\% & \cellcolor{blue!20} 69\% & 49\% & 33\% \\ 
63 & jp.gr.java\_conf.hatalab.mnv & 25\% & 24\% & 18\% & 16\% & 28\% & 19\% & \cellcolor{blue!20}  35\% \\ 
64 & org.totschnig.myexpenses & 33\% & 39\% & 24\% & 9\% & \cellcolor{blue!20} 44\% & 42\% & \cellcolor{blue!20} 44\% \\ 
65 & net.fercanet.LNM & \cellcolor{blue!20} 38\% & - & 28\% & 32\% & 36\% & 33\% & 32\% \\ 
66 & net.mandaria.tippytipper & 64\% & 46\% & 35\% & 8\% & 77\% & 47\% & \cellcolor{blue!20}  85\% \\ 
67 & es.senselesssolutions.gpl.weightchart & 47\% & 19\% & 25\% & 21\% & 39\% & \cellcolor{blue!20}  52\% & \cellcolor{blue!20}  52\% \\ 
68 & de.freewarepoint.whohasmystuff & 53\% & 53\% & 36\% & 47\% & 60\% & 56\% & \cellcolor{blue!20}  67\% \\ 
\hline
& Overall & 39\% & 31\% & 28\% & 26\% & 39\% & 35\% & \cellcolor{blue!20}  43\% \\ 
\hline
\end{tabular}}
\end{table}

\begin{figure}[tbp]
\centering
\includegraphics[width=3.5in]{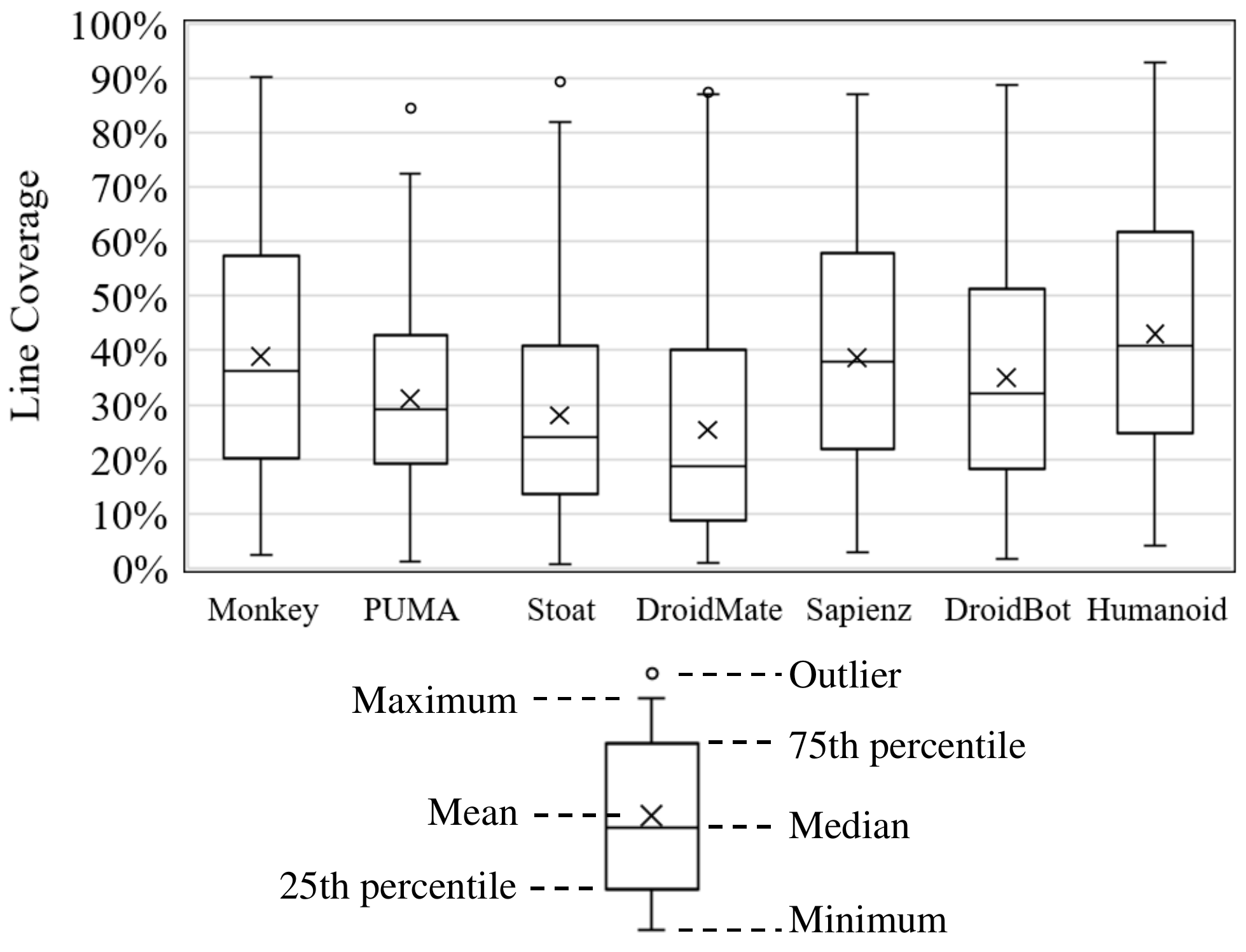}
\caption{Line coverage comparison of different testing tools over open-source apps.}
\label{figure:line_coverage_boxplot}
\end{figure}

The overall comparison of the line coverage achieved by the testing tools on open-source apps is shown in Figure \ref{figure:line_coverage_boxplot}. On average, Humanoid achieved a line coverage of 43.3\%, which is the highest across all test input generators.

It is interesting to see that Monkey, which adopts a random exploration strategy, achieved higher coverage than all other model-based testing tools except Humanoid. The fact that Monkey performs better than most other testing tools has been also confirmed by other researchers \cite{Choudhary:ASE:2015}. Because Monkey is able to generate much more inputs than other tools in the same amount of time. However, our work demonstrated the benefits of the extensibility of model-based approaches. Model-based testing tools have great potential to achieve better test performance if the GUI information is properly used.

The detailed line coverage for each app is shown in Table \ref{table:line_coverage_detailed}. For some apps such as \#3, \#36, and \#45, Monkey's coverage was significantly higher. The reason was that Monkey can generate many types of inputs (such as intents and broadcasts) that were not supported by other tools. For most of the other apps, Humanoid achieved the best results, especially for apps such as \#6, \#18, \#30.

\begin{table}[tbp]
\centering
\caption{Some app examples where Humanoid outperforms other test generators.}
\label{table:cases}
\begin{tabular}{lcc}
\hline
& \multicolumn{2}{c}{\textbf{Coverage}} \\
\textbf{App Package Name} & \textbf{Humanoid} & \textbf{Best of Others} \\ \hline
\textbf{org.passwordmaker.android}  & \textbf{58\%} & \textbf{47\%} \\
\multicolumn{3}{p{3.2in}}{- In this app, users can generate a password with a hash algorithm by selecting text, hash level, and hash method one by one. Humanoid predicted higher probabilities for these actions. Thus it was able to try more hash algorithms in a fixed length of time.} \\ \hline
\textbf{com.kvance.Nectroid}  & \textbf{53\%} & \textbf{32\%} \\
\multicolumn{3}{p{3.2in}}{- This app is a music player in which users can create a custom playlist before playing music. Humanoid completed the process of creating a playlist, while others failed to do so.} \\ \hline
\textbf{com.templaro.opsiz.aka}  & \textbf{82\%} & \textbf{61\%} \\
\multicolumn{3}{p{3.2in}}{- A core functionality of this app is converting a text message into morse code audio. To use this functionality, users have to input some text, click an \texttt{Option} button, and click a \texttt{Create} button in the order. Humanoid could generate the correct action sequence with higher probability as compared with other tools.} \\ \hline
\textbf{com.tum.yzahtzee}  & \textbf{59\%} & \textbf{44\%} \\
\multicolumn{3}{p{3.2in}}{- This app serves a number game with two text boxes and a \texttt{play} button displayed on the screen. Users need to input two numbers and then click \texttt{play} to start the game. Since Humanoid would raise the probability of inputting text and touching the \texttt{play} button, it got higher chance to start the game. Other tools kept interacting with the keyboard because it contains many clickable UI elements.} \\ \hline
\textbf{net.mandaria.tippytipper}  & \textbf{85\%} & \textbf{77\%} \\
\multicolumn{3}{p{3.2in}}{- There were many buttons on each UI state in this app. The most important ones include a small \texttt{OK} button and a \texttt{Split Bill} button. Humanoid was able to prioritize these two buttons based on the spatial information of UI.} \\ \hline
\end{tabular}
\end{table}

\begin{figure*}[h]
\centering
\includegraphics[width=7in]{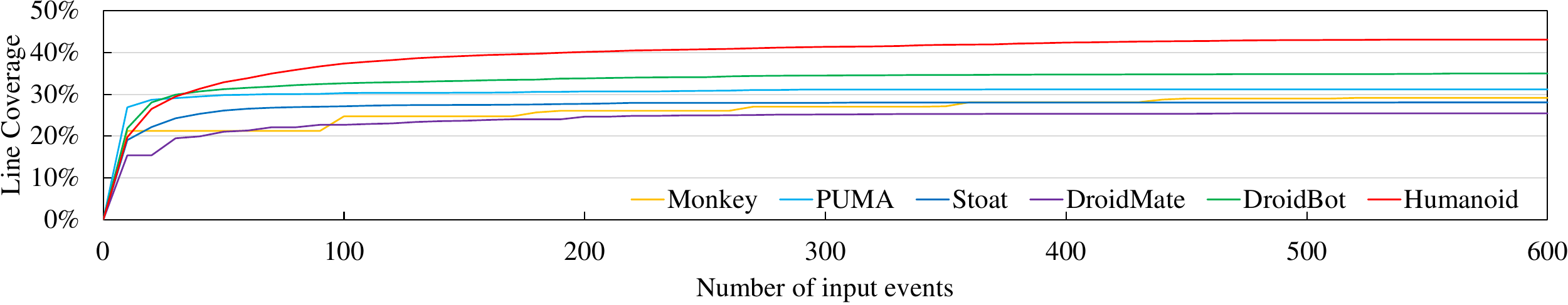}
\caption{Progressive line coverage for open-source apps.}
\label{figure:progressive_line_coverage_curve}
\end{figure*}

We further investigated why Humanoid was able to outperform other testing tools by examining the detailed test traces. We carefully inspected five apps on which Humanoid achieved significantly higher coverage. We found several cases where Humanoid behaved better than others, as illustrated in Table~\ref{table:cases}. To sum up, the high coverage of Humanoid was mainly due to two reasons: First, Humanoid was able to identify and prioritize the critical UI elements when there were plenty of UI elements to choose from. Second, Humanoid had a higher chance to perform a meaningful sequence of actions, which can drive the app into new and unexplored core functionalities.

Figure~\ref{figure:progressive_line_coverage_curve} shows the progressive coverage \textit{w.r.t.} the number of input events sent by each testing tool. Note that we did not include Sapienz in the progressive coverage figures because it sends events too fast and we could not slow it down as it was close-sourced. In the first few steps, the line coverage of all testing tools increased rapidly, as the apps were just started and all UI states were new. PUMA achieved the highest coverage in the first 10 steps because it had a strategy to restart the app at the beginning, which led to the coverage of resource recycling code in many apps. Humanoid started to lead after about 20 events. That was because the easy-reachable code was already covered at that point, and the other states were hidden behind specific interactions that can hardly be produced by other testing tools. 

At the 600th event point, the line coverage of most testing tools had almost converged, except for Monkey. This was because that the random strategy of Monkey produced a lot of ineffective and duplicated input events, which was not helpful for fast coverage improvement with regard to the number of events. However, Monkey was capable of generating much more events during the same amount of time. Its coverage continued to increase after the 600th step and finally reached about 39\% at the end of the one-hour testing duration.

\subsubsection{Testing on Market Apps}

Finally, we evaluated Humanoid with 200 popular Android apps. The experimental setup was the same as for open-source apps, however, we ran the testing process for 3 hours for each market app, as these apps are typically much larger and can reach much more UI states. Compared with the open-source apps, market apps usually have more complicated functionalities and UI structures. Thus we use this experiment to check whether Humanoid is effective for real-world scenarios. As mentioned earlier, because we do not have source code of these apps, we use activity coverage to evaluate the testing efficiency.

\begin{figure}[]
\centering
\includegraphics[width=3.2in]{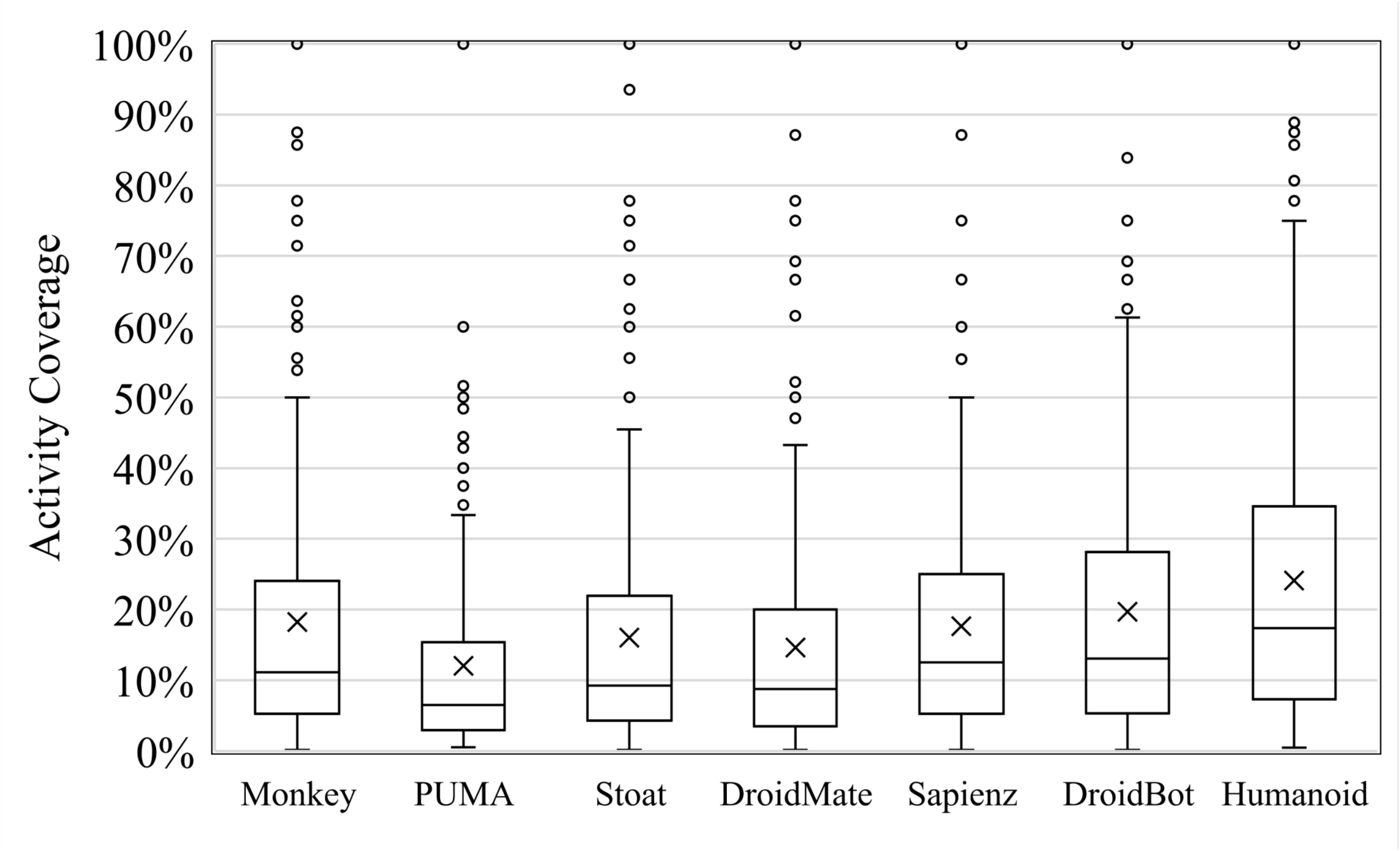}
\caption{Activity coverage comparison of the testing tools over popular market apps.}
\label{figure:activity_coverage}
\end{figure}

\begin{figure}[]
\centering
\includegraphics[width=3.0in]{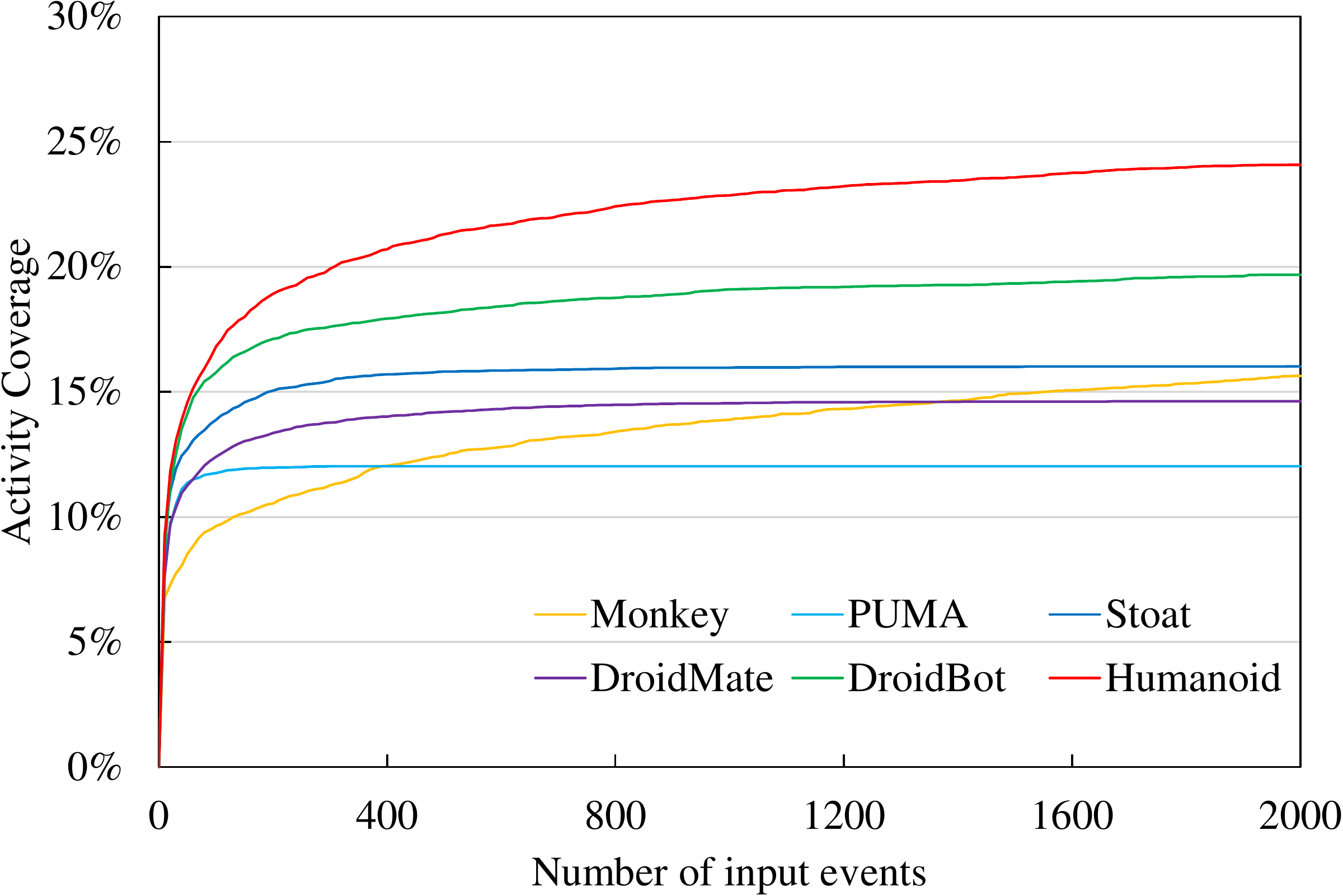}
\caption{Progressive activity coverage for market apps.}
\label{figure:progressive_activity_coverage_curve}
\end{figure}

The final activity coverage achieved by the testing tools and the progressive coverage are shown in Figure~\ref{figure:activity_coverage} and Figure~\ref{figure:progressive_activity_coverage_curve}, respectively. Similar to open-source apps, Humanoid also achieved the highest coverage (24.1\%) as compared with other tools. Due to the complexity of market apps, the coverage for some apps was not converged at the end of testing. However, we believe that Humanoid will keep the advantage even with longer testing time.

\subsection{Effectiveness of the Learned Model}
\label{eval:effectiveness}

Due to the complexity of Humanoid's input generation algorithm, it is unclear whether the high coverage of Humanoid is brought by the interaction model learned from human traces.

To check the effectiveness of the learned interaction model in improving coverage, we created a baseline algorithm by replacing Humanoid's action selection logic (line 8 in Algorithm~\ref{algorithm:strategy}) to a random policy. Then we used the baseline algorithm to test the same set of apps.

We found that the baseline algorithm without the learned interaction model almost achieved the same coverage as Humanoid, which means the human interaction model was not the main reason of coverage improvement.

\section{Limitations and Future Work}

\textbf{More types of inputs.} There are some types of inputs, such as system broadcasts and sensor events, that are not considered in this paper. This is a limitation of Humanoid because these inputs are difficult to collect from human interactions and they are also hard to represent in our deep learning model. However, it is not a severe problem as most apps can be well-tested without these actions. Humanoid also does not predict the text when sending text input actions, which can be fixed in the future by extending the interaction model or integrating other text input generation techniques \cite{textGen:ICSE:2017}.

\textbf{Further improvement in testing coverage.} Although Humanoid has been able to improve the coverage significantly from existing testing tools, the testing coverage is still far from perfect. In particular, the coverage is less than 10\% for some apps that require specific inputs such as email addresses and account/passwords. A possible solution is to design better ways of semi-automated testing, in which human testers can provide necessary guidance to the automatic tool with minimal efforts.

\textbf{Making use of textual information.} When learning the human interaction patterns, we use the UI skeleton to represent each UI state in our model, while the text in each UI element is not used. The textual information is very important for humans when using mobile apps. We believe the performance of Humanoid can be further improved if the text information can be properly represented and trained with the help of some natural language processing (NLP) techniques.

\textbf{More detailed testing on market apps.} As we do not have source code on real market apps, we cannot testing them in more details as we did for the open-source apps. However, it might be possible to reverse engineer or instrument these apps (may need to consider obfuscation as well), so that we can evaluate the testing results in more detail, for example what kind of activities are more difficult to cover, and identifying opportunities to improve Humanoid. We will consider this in our future work.



\section{Concluding Remarks}

This paper introduces Humanoid, a new GUI test input generator for Android apps that is able to predict user-preferred inputs through deep learning. Humanoid adopts a deep neural network model to learn which UI elements are more likely to be interacted by end-users and how to interact with it, from a large set of human-generated interaction traces. Experiments show that the learned model is able to accurately predict real human interaction for an Android app, but whether the learned model is effective in improving test coverage is unclear.


\bibliographystyle{IEEEtran}
\bibliography{citations}

\end{document}